\documentclass[prl,twocolumn]{revtex4-1}

\usepackage{braket}
\usepackage{mathtools}
\usepackage{amsfonts}
\usepackage[usenames,dvipsnames]{color}
\usepackage[colorlinks,bookmarks=false,citecolor=NavyBlue,linkcolor=OliveGreen,urlcolor=blue]{hyperref}

\newcommand{\be}{\begin{equation}}
\newcommand{\ee}{\end{equation}}
\newcommand{\ba}{\begin{aligned}}
\newcommand{\ea}{\end{aligned}}
\newcommand{\1}{{\rm I}}

\begin{document}

%%%%%%%%%%%%%%%%%%%%%%%%%%%%%%%%%%%%%%%%
\title{Higher-Order Hydrodynamics in 1D: a Promising Direction and a Null Result}
\author{Maurizio Fagotti}
\address{D\'epartement de Physique, \'Ecole Normale Sup\'erieure / PSL Research University, CNRS, 24 rue Lhomond, 75005 Paris, France}
%%%%%%%%%%%%%%%%%%%%%%%%%%%%%%%%%%%%%%%%

%%%%%%%%%%%%%%%%%%%%%%%%%%%%%%%%%%%%%%%%
\begin{abstract}
We derive a Moyal dynamical equation that describes exact time evolution in generic (inhomogeneous) noninteracting spin-chain models. Assuming quasistationarity, we develop a hydrodynamic theory. 
The question at hand is whether some large-time corrections are captured by  higher-order hydrodynamics. 
We consider in particular the dynamics after that two chains, prepared in different conditions, are joined together. In these situations a light cone, separating regions with macroscopically different properties,  emerges from the junction. 
In free fermionic systems some observables close to the light cone follow a universal behavior, known as Tracy-Widom scaling. 
Universality means weak dependence on  the system's details, so this is the perfect setting where hydrodynamics could emerge.
For the transverse-field Ising chain and the XX model, we show  that  hydrodynamics captures the scaling behavior close to the light cone. On the other hand, our numerical analysis suggests that hydrodynamics fails in more general models, whenever a condition is not satisfied. 
\end{abstract}

%%%%%%%%%%%%%%%%%%%%%%%%%%%%%%%%%%%%%%%%

\maketitle

%%%%%%%%%%%%%%
%\paragraph{Introduction.}  %
%%%%%%%%%%%%%%

Over the past few years, we are experiencing an increasing interest in the physics behind the nonequilibrium time evolution of inhomogeneous states. 
An example is the time evolution of  two semi-infinite chains that are joined together after having been prepared in different equilibrium conditions~\cite{BeDo16Review,VaMo16Review}. 
This kind of settings allows one to investigate the transport properties of quantum many-body systems even if the system is isolated from the environment.  

The first analytic results in this context  were obtained in noninteracting models~\cite{ARRS99,AsPi03,AsBa06,PlKa07,LaMi10,EiRz13,DVBD13,CoKa14,EiZi14,CoMa14,DeMV15,DLSB15,VSDH16,ADSV16,Bert17,Korm17,PeGa17,Platini,Ei13trapped,EiMaEv16,KoZi17,F:loco}. There, under the assumption of quasistationarity, a semiclassical picture applies where the information about the initial state is carried by free stable quasiparticles moving throughout the system. 
Similar results were obtained in the framework of conformal field theory and Luttinger liquid descriptions~\cite{BeDo12,SoCa08,CaHD08,Mint11,MiSo13,DoHB14,BeDo15,BeDo16,LLMM17,SDCV17,DuSC17,LaLeMaMo17,So16}. 
In the presence of interactions the situation was less clear~\cite{GKSS05,SaMi13,DVMR14,AlHe14,CCDH14,BDVR16,Zoto16,ViIR17,PBPD17,Ka17}, but,  eventually, Refs \cite{CaDY16,BCDF16} have shown that the continuity equations satisfied by the \mbox{(quasi)local} conserved quantities are sufficient to characterize the late-time behavior.
The framework developed in Refs \cite{CaDY16,BCDF16} is now known as \emph{generalized hydrodynamics}~\cite{CaDY16}, where ``generalized'' is used to emphasize that integrable models have infinitely many \mbox{(quasi)local} charges~\cite{Korepin}. 
We will generally omit ``generalized'' and refer to the system of equations derived in \cite{CaDY16,BCDF16} as \emph{first-order hydrodynamics}, 1$^{\rm st}$GHD, to emphasize that it is a system of first-order partial  differential equations. 
 
Within 1$^{\rm st}$GHD, it was possible to compute the profiles of local observables~\cite{BeFa16,CaDY16,BCDF16,DLCoDN17, F16,BuVaKaMo17-2,DoDuKoYo17, CoDeVi17,PiNa17}, to conjecture an expression for the time evolution of the entanglement entropy~\cite{Al17}, and to efficiently calculate Drude weights~\cite{IlNa17-0,IlNa17,DoSp17,DoSp17-DW,BuVaKaMo17}.
There are however fundamental questions that can not be addressed within 1$^{\rm st}$GHD; diffusive transport~\cite{Sp17,MeKaPr17,MeKlPr17,LjZnPr17,LjZnPr1t_NC,MiMaKr17} and large-time corrections~\cite{EiRz13,Bert17,Korm17,PeGa17} are two of them. The importance of these issues results in a considerable urge to fill these gaps~\cite{DoSp17},  passing through refinements and reinterpretations of the theory~\cite{Bu17,DoYo16,DoSpYo17,DoToCa17,PiNa17}.

In these notes we carry out a preliminary analysis of  whether higher-order hydrodynamics gives access to additional physical information. 
Since any refinement to the equations governing the dynamics must be able to pass the noninteracting test, we focus on generic noninteracting spin-chain models. We develop a complete hydrodynamic theory, GHD, based on the single assumption of quasistationarity. Within GHD, we compute some large-time corrections and compare them with exact numerical data. The result is puzzling: higher-order hydrodynamics reproduces  known asymptotic behaviors  in the XX model and in the transverse-field Ising chain; the same hydrodynamic description, however, seems to fail in generic noninteracting models. 
  
%%%%%%%%%%%%%%
\paragraph{The system.}  %
%%%%%%%%%%%%%%

We consider an infinite spin-$\frac{1}{2}$ chain described by a Hamiltonian of the form
\be\label{eq:H}
\boldsymbol H=\sum_{\ell\in \mathbb Z}\sum_{n\in\mathbb N_0}\sum_{\alpha,\beta\in x,y}J_{\ell,n}^{\alpha,\beta} \boldsymbol \sigma_\ell^\alpha \boldsymbol \Pi_{\ell,n}^z\boldsymbol \sigma_{\ell+n}^\beta+\sum_{\ell\in\mathbb Z} J_\ell^z \boldsymbol\sigma_\ell^z\, ,
\ee
where $\boldsymbol \sigma_\ell^\alpha$ are Pauli matrices, $\boldsymbol\Pi_{\ell,n}^z=\prod_{j=\ell+1}^{\ell+n-1}\boldsymbol\sigma_j^z$ ($\boldsymbol\Pi_{\ell,0}^z$ is the identity $\boldsymbol\1$), $\mathbb Z$ is the set of all the integers, and $\mathbb N_0$ is its nonnegative subset.
Under the Jordan Wigner transformation $\boldsymbol a_{2\ell-1}=\prod_{j<\ell}\boldsymbol \sigma_j^z \boldsymbol \sigma_\ell^x$, $\boldsymbol a_{2\ell}=\prod_{j<\ell}\boldsymbol \sigma_j^z \boldsymbol \sigma_\ell^y$, the Hamiltonian is mapped into a chain of noninteracting Majorana fermions ($\{\boldsymbol a_\ell,\boldsymbol a_n\}=2\delta_{\ell n}\boldsymbol\1$, where $\{\cdot,\cdot\}$ is the anticommutator, and $\delta_{\ell n}$ is the Kronecker delta)
\be
\boldsymbol H=\frac{1}{4}\sum\nolimits_{\ell, n\in\mathbb Z}\boldsymbol a_\ell \mathcal H_{\ell n} \boldsymbol a_n\, .
\ee 
Here $\mathcal H$ is an infinite~\cite{f:3} purely imaginary antisymmetric matrix.  
This class includes several paradigmatic models, as the transverse-field Ising chain~\cite{Pfeuty} and the XY model~\cite{LSM:61}. 
Being quadratic, $\boldsymbol H$ is diagonal in a basis of Slater determinants. These are states $\ket{\Gamma}$ completely characterized (up to a phase) by the fermionic two-point functions, which can be organized in a purely imaginary  antisymmetric matrix $\Gamma$, known as ``correlation matrix''
\be\label{eq:Gamma}
\Gamma_{\ell n}=\delta_{\ell n}-\braket{\Gamma|\boldsymbol a_\ell \boldsymbol a_n|\Gamma}\, .
\ee 
Thermal states are Slater determinants as well;
the ground state, however, is not always a Slater determinant, as a symmetry could be spontaneously broken. 

%%%%%%%%%%%%%%%%%%%%%%%%%%%
\paragraph{The equations governing the dynamics.} %
%%%%%%%%%%%%%%%%%%%%%%%%%%%

Quadratic operators are closed under commutation, so a Slater determinant that time evolves under a Hamiltonian of the form \eqref{eq:H} remains a Slater determinant. Specifically, the time evolving state is as follows
\be\label{eq:evol}
e^{-i \boldsymbol H t}\ket{\Gamma}= e^{-i\gamma_t}\ket{e^{-i\mathcal H t}\Gamma e^{i\mathcal H t}}\, ,
\ee
where $e^{-i\gamma_t}$ is a phase.
Let $\mathcal M$ be an infinite purely imaginary antisymmetric matrix with elements decaying sufficiently fast to zero  the farther they are from the main diagonal.
We define the $2\kappa$-by-$2\kappa$ symbol $\hat m_x(e^{i p})$ of  $\mathcal M$ as follows
\be\label{eq:symbol}
\mathcal M_{2\kappa\ell_\kappa+i,2\kappa n_\kappa+j}=\!\!\int_{-\pi}^{\pi}\!\!\frac{\mathrm d p}{2\pi}e^{i(\ell_\kappa-n_\kappa) p}[\hat m_{\frac{\ell_\kappa+n_\kappa}{2}}(e^{i  p})]_{i j} \, ,
\ee
where $i,j=1,\dots,2\kappa$. 
The symbol $\hat m_{x}(e^{i  p})$ enters this equation only with $x\in\frac{1}{2}\mathbb Z$, where $\frac{1}{2}\mathbb Z$ is the set of all the integers and the half-integers. In addition,  if $x$ is \mbox{(half-)integer}, the equation only fixes the \mbox{$\pi$-(anti-)periodic} part of $\hat m_{x}(e^{i  p})$ with respect to $p$. The undefined parts of the symbol are irrelevant and can be chosen arbitrarily. It is convenient to require the  symbol to be Hermitian and to satisfy $[\hat m_{x}(e^{i p})]^t=-\hat m_{x}(e^{-i p})$, where ${}^t$ denotes transposition. 
We can then extend its definition in a smooth way so as to allow for real $x\in\mathbb R$. 

We find that the degrees of freedom in the definition of the symbol can be used to recast time evolution \eqref{eq:evol} in the form of a Moyal dynamical equation~\cite{LONG}
\be\label{eq:fond}
i\partial_t \hat \Gamma_x(e^{ip})=\hat h_{x}(e^{i p})\star\hat\Gamma_{x}(e^{i p})-\hat \Gamma_{x}(e^{i  p})\star\hat h_{x}(e^{i  p})\, .
\ee
Here $\hat \Gamma_x(e^{ip})$ is the symbol of the correlation matrix, $\hat h_{x}(e^{i p})$ is the symbol of the Hamiltonian, and $\star$ denotes the Moyal star product~\cite{Moyal}, defined as
$
\hat f_x(e^{ip})\star \hat g_x(e^{ip})=e^{i\frac{\partial_{q}\partial_x-\partial_{p}\partial_y}{2}} \hat f_x(e^{ip})\hat g_y(e^{iq})|_{q=p\atop y=x}
$.
We note that connections between matrix multiplication and Moyal star product have been already established  (see, \emph{e.g.}, Ref.~\cite{Hansen}). 
Our particular mapping, based on \eqref{eq:symbol},  allows for a simple interpretation of the various quantities. 
For example, if the Hamiltonian is invariant under a shift by $\kappa$ sites, we can impose $\hat h_{x}(e^{i  p})=\hat h(e^{i  p})$, and  the excitation energies $\varepsilon_n(p)$ are the eigenvalues of  $\hat h(e^{i p})$, in the sense that $\hat h(e^{i p})=\sum_{n=1}^\kappa \varepsilon_n(p)\mathrm{P}_n(p)-\varepsilon_n(-p)\mathrm{P}^t_n(-p)$, where $\mathrm{P}_n(p)$ and $\mathrm{P}^t_n(-p)$ are projectors orthogonal to one another. 
In that case,
\eqref{eq:fond} can be solved; its solution reads
\begin{multline}\label{eq:exactsolhom}
\hat \Gamma_{x,t}(e^{i p})=\iint_{-\infty}^\infty\frac{\mathrm d y  \mathrm d q}{2\pi} e^{i q (x-y)}\\
e^{-i t \hat h(e^{i (p+\frac{q}{2})})}\hat \Gamma_{y,0}(e^{i p})e^{i t \hat h(e^{i (p-\frac{q}{2})})}\, .
\end{multline}
This can be interpreted as an exact Wigner description~\cite{Wigner} of the dynamics in noninteracting spin-chain models. The solution \eqref{eq:exactsolhom} applies to any Slater determinant time evolving under any homogeneous noninteracting Hamiltonian. 
Eqs \eqref{eq:fond} and \eqref{eq:exactsolhom} do not seem to be widely known, but their structure can be recognized in equations emerging within semiclassical approximations, as in Ref.~\cite{BaDL17}. 

%we are not aware of any work where \eqref{eq:fond} or \eqref{eq:exactsolhom} have been used to investigate the dynamics in spin-chain models, suggesting that  they are not widely known. 

%%%%%%%%%%%%%%%%
\paragraph{Hydrodynamics.}  %
%%%%%%%%%%%%%%%%

Eq.~\eqref{eq:fond} is as general as it is exceptional, applying only to noninteracting models. It is a useful tool, but it can not be easily generalized in the presence of interactions. In addition, even when \eqref{eq:exactsolhom} applies, the explicit calculation of the integrals, but also their numerical evaluation, can be difficult. Notwithstanding, one is often interested in particular aspects of the dynamics that are not expected to depend on all the system's details. For these two reasons, we pivot to a description that, a priori, is only an approximation; we develop a hydrodynamic theory. 

To that aim, we add the hypothesis of \emph{quasistationarity}. In other words, we assume that $\ket{\Gamma}^{\rm hyd}$  is a locally quasistationary state~\cite{BeFa16} at every time. For homogeneous Hamiltonians, this condition is equivalent to ask for the symbol of the correlation matrix to locally commute with the symbol of the Hamiltonian, \emph{i.e.} $[\hat\Gamma^{\rm hyd}_{x,t}(e^{ip}),\hat h(e^{ip})]=0$. If we extract the diagonal part of \eqref{eq:fond} in a basis that diagonalizes  $\hat h(e^{ip})$ and replace $\hat\Gamma_{x,t}(e^{ip})$ by $\hat\Gamma^{\rm hyd}_{x,t}(e^{ip})$, we find 
\begin{multline}\label{eq:hydro}
i\partial_t\hat \Gamma^{\rm hyd}_{x,t}(e^{i p})=\iint_{-\infty}^\infty\frac{\mathrm d q\mathrm d y}{2\pi} e^{i q (x-y)}\\
\langle \langle\hat h(e^{i  (p+\frac{q}{2})}) -\hat h(e^{i  (p-\frac{q}{2})})\rangle \rangle(e^{i p}) \hat\Gamma^{\rm hyd}_{y,t}(e^{i p})\, ,
\end{multline}
where $\langle \langle \hat a(e^{i k})\rangle\rangle(e^{ip})$  denotes the diagonal part of $\hat a(e^{i k})$ in a basis that diagonalizes $\hat h(e^{i p})$.
This is a \emph{complete hydrodynamic equation}. 
 In fact, \eqref{eq:hydro} can be put in a more familiar form if $\kappa$ is so large that the Hamiltonian has only nearest neighbor couplings. In that case, \eqref{eq:hydro} reads~\cite{f:4}
\be\label{eq:hydrostandard}
\partial_t\rho^{{\rm hyd}}_{n;x,t}(p)+ v_n(p)[\rho^{{\rm hyd}}_{n;x+\frac{1}{2},t}(p)-\rho^{{\rm hyd}}_{n;x-\frac{1}{2},t}(p)]=0\, ,
\ee
where $n=1,\dots,\kappa$,  $\rho^{{\rm hyd}}_{n;x}(p)$ are the so-called  root densitied, and $v_n(p)$ are the velocities of the excitations. Roughly speaking \cite{f:1}, $\rho^{{\rm hyd}}_{n;x}(p)$ describes the density of the $n$-th species of excitations over the ground state~\cite{f:0}; up to  additive constants and multiplicative factors, the root densities are the eigenvalues of $\hat \Gamma^{\rm hyd}_{x,t}(e^{i p})$. 
 The first-order hydrodynamic equation is recovered in the limit of weak inhomogeneity, which allows one to expand the last two terms of \eqref{eq:hydrostandard} about $x$, ignoring the contributions from spatial derivatives higher than the first. That is the equation that Refs~\cite{CaDY16,BCDF16} generalized to interacting integrable models. Incidentally, applying it to \eqref{eq:hydrostandard} the same prescription that lifted its first-order approximation to a theory for interacting integrable models results in
\be\label{eq:hydrogen}
\partial_t\rho^{{\rm hyd}}_{n;x,t}+ v^{\rm hyd}_{n;x+\frac{1}{2}}\rho^{{\rm hyd}}_{n;x+\frac{1}{2},t}-v^{\rm hyd}_{n;x-\frac{1}{2}}\rho^{{\rm hyd}}_{n;x-\frac{1}{2},t}=0\, .
\ee
Here the velocity depends on $x$ (and, in turn, it is affected by the assumption of quasistationarity) because it is dressed by the interaction~\cite{BpEsLa14}. 

At this stage, \eqref{eq:hydrogen} is nothing but  a provocation; 
we must  first understand what physical information is contained in  \eqref{eq:hydrostandard} in the very noninteracting case. Indeed, \eqref{eq:hydrostandard} is based on the assumption of quasistationarity, which is known to be exact only in particular limits when \eqref{eq:hydrostandard} reduces to its first-order approximation~\cite{LONG}.

%%%%%%%%%%%%%%%%%%%%%%%%
\paragraph{Dynamics close to the light cone.} %
%%%%%%%%%%%%%%%%%%%%%%%%

Let us imagine to prepare two semi-infinite chains in different stationary conditions, for example at different temperatures. Let us then join the chains together so as to form a single infinite chain. The state is let to evolve under the merged Hamiltonian, which we assume to be homogeneous.  
Qualitatively, a light cone, which separates regions with macroscopically different properties, emerges from the junction.  1$^{\rm st}$GHD turns out to capture the limit of large time at any position. 
On the other hand, the large-time corrections are generally beyond its capabilities. 

We wonder whether some corrections can still be computed  within GHD. Since off-diagonal contributions are neglected, it is not reasonable to expect hydrodynamics to capture generic corrections. 
It could be effective, however, for corrections  exhibiting universal properties, like the ones studied in Ref.~\cite{EiRz13}. 
There, the time evolution of a domain-wall state under a free fermionic Hamiltonian (XX model) was considered.
The authors were able to establish a connection  between the probability distributions of particular  observables close to the light cone  and the distribution functions of the largest eigenvalues of the Gaussian unitary random matrix ensemble~\cite{AiryK}. 
In particular, the two-point functions of the fermions lying in a region, around the edge, scaling as $t^{\frac{1}{3}}$ have corrections that decay as $t^{-\frac{1}{3}}$, and they can be written in terms of the so-called Airy kernel 
\be\label{eq:Airykernel}
K_1(u,v)=({\rm Ai}(u){\rm Ai}'(v)-{\rm Ai}(v){\rm Ai}'(u))/(u-v)\, .
\ee
An analogous behavior was observed years before~\cite{Platini}, studying the transverse-field Ising chain, and, recently, Refs~\cite{PeGa17,Korm17} pointed out that, also in that case, the large-time corrections are characterized by the Airy kernel. 

The presumptive universality of these corrections makes them a perfect candidate to test GHD. 
As a first step, we argue that, if the scaling behavior close to the light cone can be described by complete hydrodynamics, then it can also be described by  
hydrodynamics at the third order. The latter is obtained by expanding at the third order, either the integrand in \eqref{eq:hydro} about $p$, as if $q$ were close to zero, or, equivalently, the last two terms of \eqref{eq:hydrostandard} about $x$. 
For the sake of simplicity, we restrict ourselves to one-site shift invariant  Hamiltonians ($\kappa=1$). 
Third-order hydrodynamics reads
\be\label{eq:hydro3}
\partial_t\rho^{{\rm hyd}_3}_{x,t}(p)+ v(p)\partial_x\rho^{{\rm hyd}_3}_{x,t}(p)+\frac{w(p)}{24}\partial_x^3 \rho^{{\rm hyd}_3}_{x,t}(p)=0\, ,
\ee
where $w(p)$ is a function that, generically, can not be written only in terms of the dispersion relation; its expression is reported in Ref.~\cite{LONG}. The reader seeking for a concrete example can assume that the Hamiltonian has nearest neighbor couplings, so that $w(p)=v(p)$ (see \eqref{eq:hydrostandard}). 
The solution to \eqref{eq:hydro3} is 
\be
\label{eq:hydro3sol}
\rho^{\rm hyd_3}_{x,t}(p)=\int_{-\infty}^{\infty}\mathrm d y\mathrm{Ai}(y) \rho^{\rm hyd_3}_{x-v(p) t-\frac{y}{2}\sqrt[3]{w(p) t} ,0}(p)\, ,
\ee
where $\mathrm{Ai}(y)$ is the Airy function. 
We focus on the situation where $v(p)$ has a unique global maximum at $p=\bar p$ and $v^{\prime\prime}(\bar p)$ is nonzero. 
The maximal velocity $v(\bar p)$ determines the speed at which the light cone propagates to the right. 
We now sketch a proof of the equivalence between \eqref{eq:hydro3sol} and complete hydrodynamics close to the right light cone. 
It is convenient to start from \eqref{eq:hydrostandard}. For $x\in\frac{1}{2}\mathbb Z$ (these are the only relevant positions, \emph{cf}. \eqref{eq:symbol}), its solution reads
\be\label{eq:gensolhydro}
\rho^{{\rm hyd}}_{n;x,t}(p)=\rho_n^{+}(p)+[\rho_n^{-}(p)-\rho_n^{+}(p)]\sum_{\nu=2x}^\infty\!\! J_\nu(2v_n(p)t)\, ,
\ee
where $\rho_n^{-}$ and $\rho_n^{+}$ are the root densities describing the initial state on the left ($x\leq 1$) and on the right hand side ($x>0$) of the junction, respectively~\cite{f:2}; $J_\nu(z)$ is the Bessel function of the first kind. 
Generally, close to the right light cone, a single species of excitations has a root density significantly different from the corresponding $\rho_n^+(p)$; let us focus on that species. It turns out that the leading contribution in the sum of \eqref{eq:gensolhydro} is carried by the Bessel functions with order $\nu$ and argument $z$ sufficiently close to $2x$. Each Bessel function can then be replaced by the leading term in its uniform asymptotic expansion,  Eq. (9.3.23) of Ref.~\cite{AbSt}. In the limit of large time, the terms turn out to be smooth functions of $\nu$, so the sum in \eqref{eq:gensolhydro} can be replaced by an integral. We finally obtain \eqref{eq:hydro3sol}, proving that  third-order hydrodynamics is equivalent to complete hydrodynamics~\cite{LONG}.

 By Wick's theorem, the expectation values of any observable can be written in terms of the correlation matrix~\eqref{eq:Gamma}; we can focus on its time evolution. 
In the hypothesis of quasistationarity, $\Gamma$ reads (see also Ref.~\cite{F16})
\begin{multline}\label{eq:Gammahyd3}
\Gamma_{2\ell+i, 2m+j}(t)=2\int_{-\pi}^\pi\mathrm d p \Bigl(\rho^{\rm hyd}_{\frac{\ell+m}{2},t}(p)-\frac{1}{4\pi}\Bigr)\times\\
\Bigl[\cos((\ell-m)p)A_{i, j}(p)+i \sin((\ell-m)p) B_{i, j}(p)\Bigr]\, ,
\end{multline}
where $A(p)=\frac{\tilde \sigma(p)+\tilde \sigma(-p)}{2}$ and $B(p)=\1+\frac{\tilde \sigma(p)-\tilde \sigma(-p)}{2}$, 
with $\tilde\sigma(p)=\mathrm{sgn}\{\hat h(e^{i p})-\frac{\1}{2}\mathrm{tr}[\hat h(e^{i p})]\}$. We note that the dispersion relation is $\varepsilon(p)=\mathrm{tr}[\frac{\1+\tilde\sigma(p)}{2}\hat h(e^{i p})]$ and, as usual, the velocity is defined as $v(p)=\varepsilon'(p)$.
We consider observables that lie close to the light cone. 
These observables can be fully described by the reduced density matrix of a spin block consisting of the sites $S=\{r,r+1,\dots,r+|S|-1\}$ lying around the edge, that is to say $\frac{r}{t}\rightarrow  v(\bar p)$ and $|S|\ll v(\bar p)t$. 
The reduced density matrix is a Slater determinant as well, with the following block correlation matrix \be
[\Gamma^{(S)}_{\ell,m}]_{i, j}(t)=\Gamma_{2(\ell+r)+i, 2(m+r)+j}(t)\, ,
\ee 
where $\ell,m=0,\dots,|S|-1$ and $i,j=1,2$. Following Ref.~\cite{EiRz13}, we consider the scaling limit where the time is large and both $r-v(\bar p)t$ and the subsystem's length $|S|$ are proportional to~$ t^{\frac{1}{3}}$. Assuming the various functions to be sufficiently smooth around $p=\bar p$ we find
\begin{multline}\label{eq:Gammasol}
\Gamma^{(S)}_{\ell,m}(t)\approx \Gamma^{(S)}_{\ell,m}(0)+4\pi \Bigl(\frac{2\alpha}{-\bar v^{\prime\prime}t}\Bigr)^{\frac{1}{3}}
[\bar \rho^{-}-\bar \rho^{+} ]K_\alpha(x_\ell,x_m)\times \\
[\cos((\ell-m)\bar p)\bar A+i \sin((\ell-m)\bar p) \bar B]\, ,
\end{multline}
where   $\bar f$ stands for $f(\bar p)$; $\alpha=\mathrm{sgn}(\bar w)(\frac{|\bar w|}{-\bar v^{\prime\prime}})^{\frac{1}{2}}$, and we introduced the rescaled variables
\be\label{eq:rescaled}
x_j=2^{\frac{1}{3}}\frac{j+r-\bar v t}{\sqrt[3]{-\bar v^{\prime\prime}\alpha^2 t}}\, ;
\ee 
the kernel $K_{\alpha}(u,v)$ is defined as follows
\begin{multline}
K_\alpha(u,v)=2^{\frac{2}{3}}\int_{0}^\infty\mathrm d y \mathrm{Ai}\bigl[\mathrm{sgn}(\alpha) 2^{\frac{2}{3}}\bigl(y+\frac{u+v}{2}\bigr)\bigr]\times\\
\sin[\alpha (u-v)\sqrt{y}]/[\alpha\pi(u-v)]\, .
\end{multline}
Using a representation of the product of two Airy functions derived in Ref.~\cite{Airy}, it is simple to show that, if $\alpha>0$, $K_\alpha(u,v)$ can be expressed in terms of the Airy kernel \eqref{eq:Airykernel}, as $K_\alpha(u,v)=K_1(\frac{1+\alpha}{2}u+\frac{1-\alpha}{2}v,\frac{1+\alpha}{2}v+\frac{1-\alpha}{2}u)$.
In the transverse-field Ising chain and in the XX model, the parameter $\alpha$ is equal to unity, and we recover \eqref{eq:Airykernel}. 

\begin{figure}[t]
\includegraphics[width=0.43\textwidth]{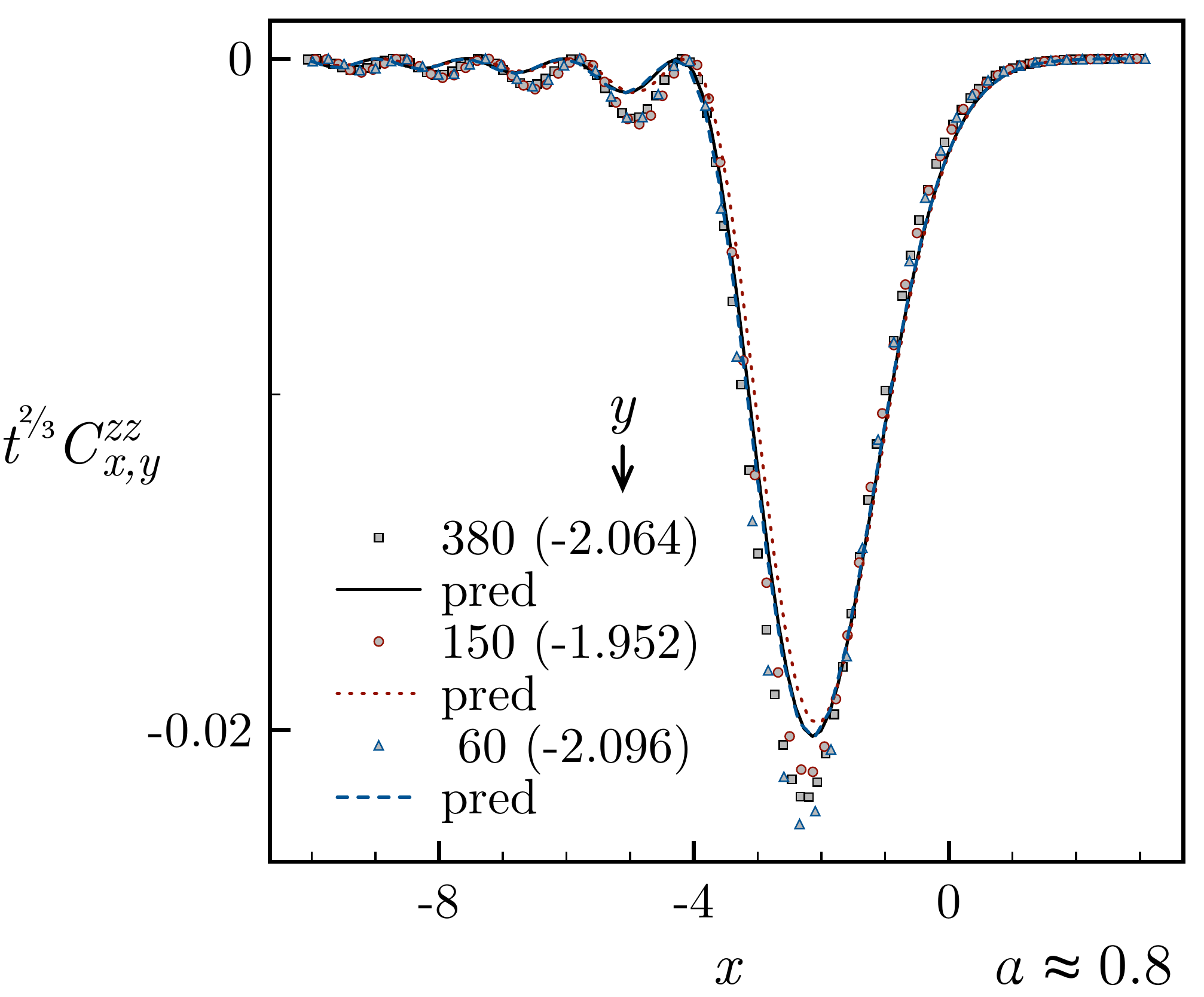}
\begin{center}
\caption{The connected two-point function of $\boldsymbol s^{z}$ as a function of the rescaled positions   $x,y$ \eqref{eq:rescaled} in the XY model with $J_{\ell,0}^{xx}=-1.15$, $J_{\ell,0}^{yy}=0.15$, and $J_{\ell}^{z}=-2$ (see \eqref{eq:H}) ($\alpha\approx 0.8$). The symbols are exact numerical data (in a chain with $1801$ spins) at three different times $t=60,150,380$, after that a thermal state with inverse temperature $\beta=1$ has been put in contact with the infinite-temperature state. The lines are the predictions \eqref{eq:Czz}.
The agreement is fair, but the data do not seem to approach the predictions as the time is increased. 
}\label{fig:1}
\end{center}
\end{figure}

A comparison with Ref.~\cite{Korm17} shows that, in  the transverse-field Ising chain, GHD gives the correct asymptotic behavior.  
The same conclusion can be drawn for the XX model. More generally,  whenever $\alpha=1$, GHD seems to capture the behavior close to the light cone. 

It is worth noting that \eqref{eq:Gammasol}  describes the asymptotic behavior only if, (i), the
difference of the root densities at the initial time is nonzero at $p=\bar p$, 
and, (ii), the expectation value of the observable does not accidentally zero the term. The former case is closely related to the situation studied in Ref.~\cite{CoDeVi17} for a similar protocol in the XXZ model; the latter case is discussed in Ref.~\cite{PeGa17} considering the critical Ising model.  

We analyze the behavior for $\alpha\neq 1$ numerically. 
Fig.~\eqref{fig:1} shows the edge profile of the connected two-point function of $\boldsymbol s^z=\frac{1}{2}\boldsymbol \sigma^z$, \emph{i.e.}, $C^{zz}_{x_\ell,x_m}=\braket{\boldsymbol s_{\ell+1}^z\boldsymbol s_{m+1}^z}-\braket{\boldsymbol s_{\ell+1}^z}\braket{\boldsymbol s_{m+1}^z}$, after putting in contact a thermal state at inverse temperature $\beta $ with the infinite temperature state. In that case we have  ($x\neq y$)
\be\label{eq:Czz}
C^{zz}_{x,y}\approx-4\pi^2 \cos^2\bar \theta\tanh^2(\frac{\beta\bar \varepsilon}{2}) K_\alpha^2(x,y)\sqrt[3]{\frac{4\alpha^2}{(\bar v^{\prime\prime})^2 t^2}}
\, .
\ee
We report data for the XY model, which is described by a Hamiltonian of the form \eqref{eq:H} with the only nonzero coupling constants $J_{\ell,0}^{xx}$, $J_{\ell,0}^{yy}$, and $J_\ell^z$. The predictions are only in a fair agreement with the numerical data, and we do not see a substantial reduction of the discrepancy when the time is increased.  
It is still possible that the times considered are not sufficiently large, but we find it more likely that, for $\alpha\neq 1$, GHD is not exact. 

%%%%%%%%%%%%%%%%%%%%%
\paragraph{Summary and discussion. } %
%%%%%%%%%%%%%%%%%%%%%

We have derived a Moyal dynamical equation that  describes exact time evolution in noninteracting spin-chain models. Assuming quasistationarity, we developed a hydrodynamic theory. We identified the neighborhood of the light cone (emerging from the junction of two steady states) as a region where higher-order hydrodynamics could improve on 1$^{\rm st}$GHD.  
In the XX model and in transverse-field Ising chain, our expectations are met. In more general systems, we report a discrepancy every time that the condition $\alpha=1$ is not satisfied. 
In models with nearest-neighbor couplings, $\alpha\neq 1$ is equivalent to $v(\bar p)\neq - v^{\prime\prime}(\bar p)$, where $v(p)$ is the velocity and $\bar p$ is the momentum at which the velocity is maximal. At the moment, we do not see why this condition should matter.  
A definite answer can be given working out the exact solution~\eqref{eq:exactsolhom}. That is a cumbersome calculation that we leave to future works. 

%We point out that, even if \eqref{eq:Gammasol} does not include the whole correction,  it it is not necessarily spurious, \emph{i.e.}, it could capture particular contributions in the asymptotic expansion of \eqref{eq:exactsolhom}.

%but we find it more likely that, for $\alpha\neq 1$, hydrodynamics is not exact.
%
%This result casts doubts on the possibility to apply \eqref{eq:hydrogen}, or, more generally, to keep the assumption of quasistationarity beyond first-order hydrodynamics. 

%%%%%%%%%%%%%%
%\paragraph{Conclusions.} %
%%%%%%%%%%%%%%

%We have derived a Moyal dynamical equation that  describes exact time evolution in noninteracting spin-chain models. Assuming quasistationarity, we developed a hydrodynamic theory. 
%We focussed on the large-time corrections close the light cone emerging from the junction of two steady states.
%We showed that third-order hydrodynamics is equivalent to complete hydrodynamics. In that framework, we computed the scaling behavior of the observables close to the light cone. For the XX model and the transverse-field Ising chain, our predictions are in agreement with known exact results. Nevertheless, there are numerical indications that,  in more general systems, higher-order hydrodynamics fails.

%%%%%%%%%%%%%%
\begin{acknowledgments} %
%%%%%%%%%%%%%%
In \cite{SM}, the reader can find numerical evidence of the validity of the equations derived.
I thank Andrea De Luca and Pierre Le Doussal for useful discussions. 

I  acknowledge support by LabEx ENS-ICFP:ANR-10-LABX-0010/ANR-10-IDEX-0001-02 PSL*.

%%%%%%%%%%%%%
\end{acknowledgments}%
%%%%%%%%%%%%%

%%%%%%%%%%%%%%%%%%%%%%%%%%%%%%%%%%%%%%%%%%

\onecolumngrid

\newpage
%%%%%%%%%%%%%%%%%%%%%%%%%%%%%
\begin{center}
{\large{\bf Higher-Order Hydrodynamics in 1D: a Promising Direction and a Null Result (Supplemental Material)}}
\end{center}
%%%%%%%%%%%%%%%%%%%%%%%%%%%%%
We refer the reader to \cite{LONG} for more details on the derivation of the results presented in the main text. 
This supplemental material only aims at providing some numerical evidence of the validity of the equations. 

\begin{figure}[h]
\includegraphics[width=0.4\textwidth]{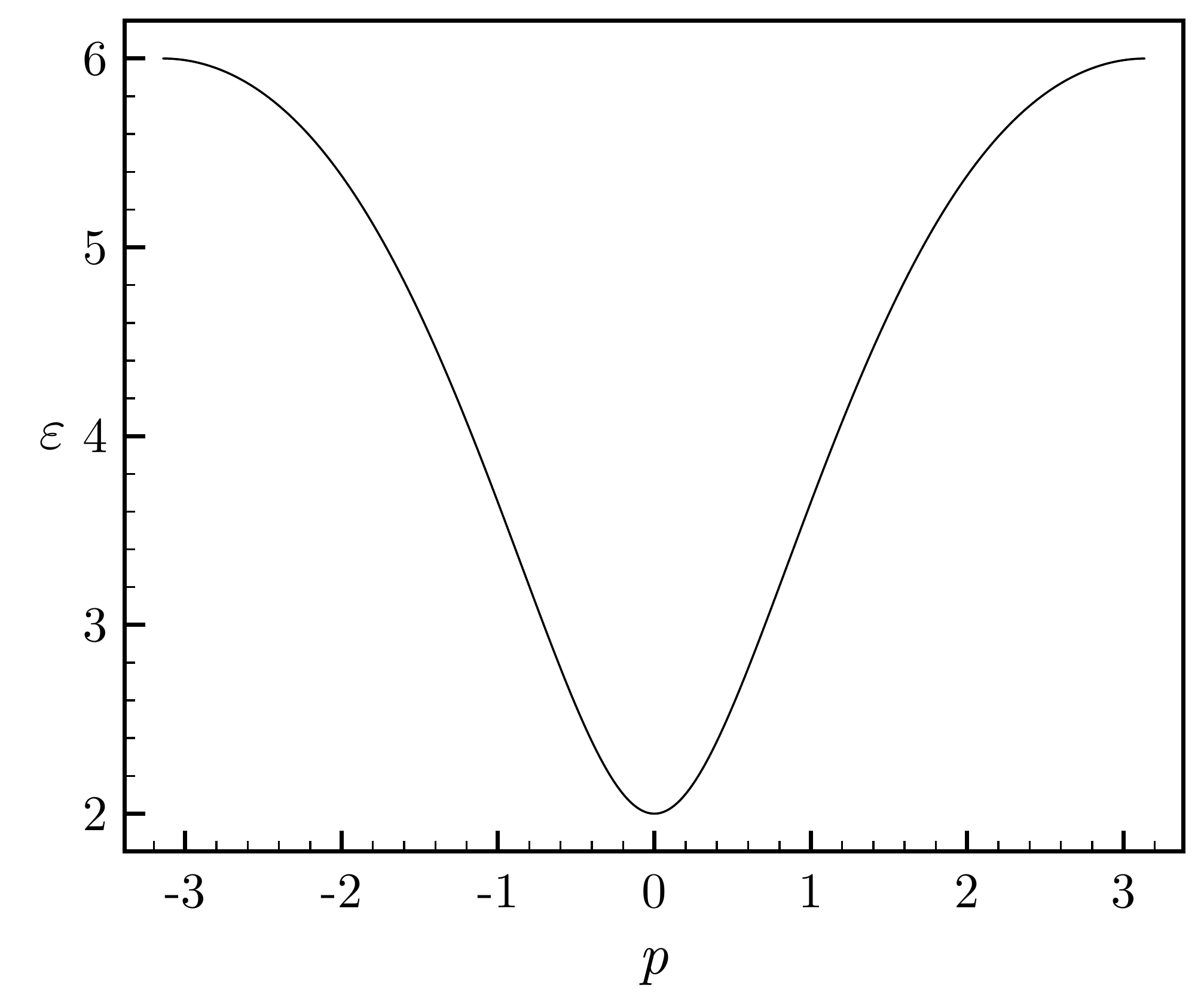}\qquad\qquad
\includegraphics[width=0.4\textwidth]{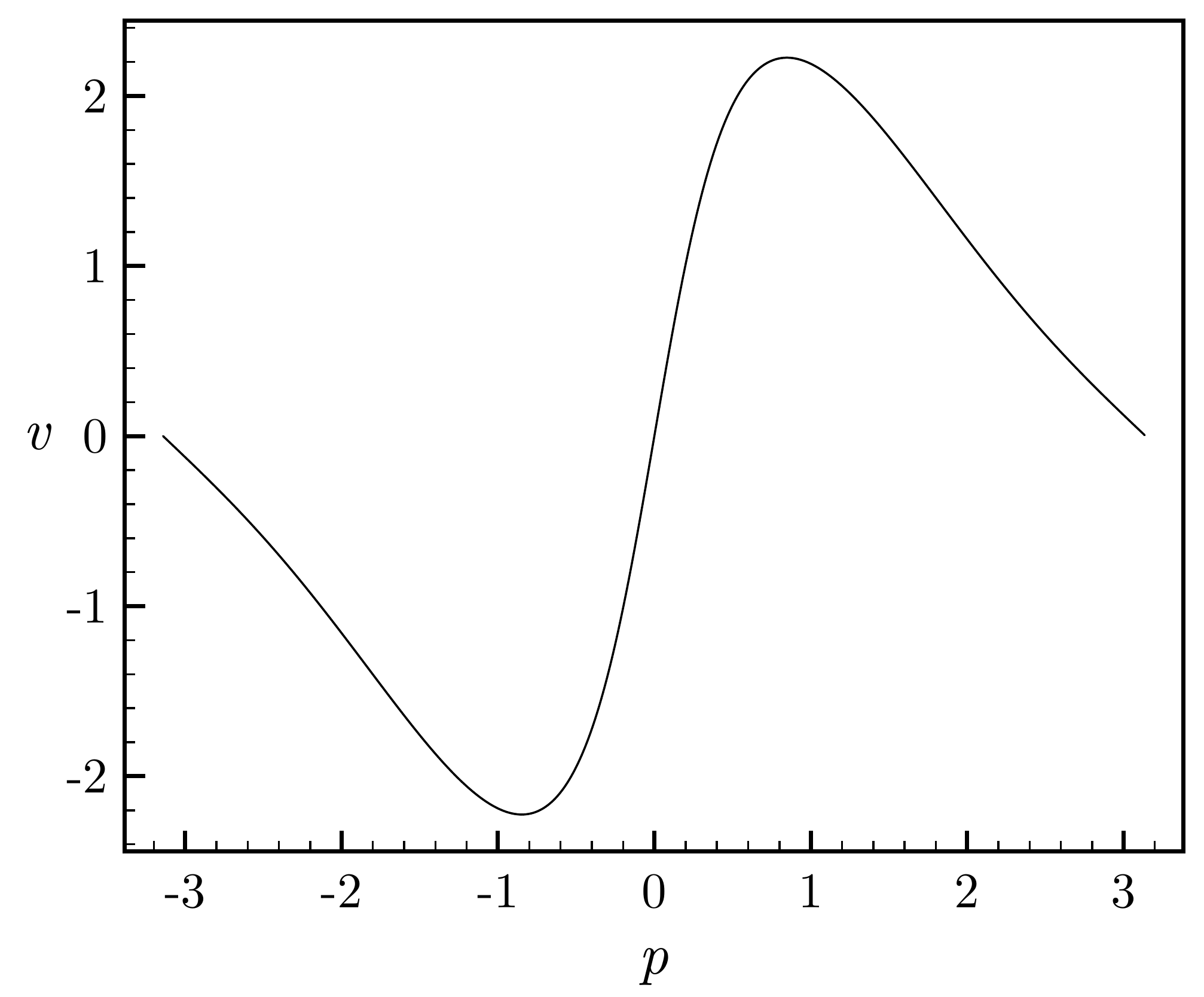}
\begin{center}
\caption{The dispersion relation (left panel) and the velocity (right panel) of the excitations for the Hamiltonian \eqref{SM:HXY}.  
}\label{fig:eps}
\end{center}
\end{figure}

Since we reported an unexpected discrepancy between the hydrodynamic predictions and the exact numerical data, we consider the same Hamiltonian as in the example of Fig.~\ref{fig:1}, specifically
\be\label{SM:HXY}
\boldsymbol H=-\sum_\ell \Bigl(1.15 \boldsymbol \sigma_\ell^x\boldsymbol \sigma_{\ell+1}^x-0.15\boldsymbol \sigma_\ell^y\boldsymbol \sigma_{\ell+1}^y+2\boldsymbol \sigma_\ell^z\Bigr)\, .
\ee
For the reader's convenience, the dispersion relation $\varepsilon(p)$ and the velocity $v(p)$ are shown in Fig.~\ref{fig:eps}. The Hamiltonian parameters have been chosen in such a way that $\varepsilon(p)$  and $v(p)$ do not differ much from those in the transverse-field Ising chain. 
\begin{description}
\item[Check of the Moyal dynamical equation]
Here we check the validity of \eqref{eq:exactsolhom}. 
We choose an initial state 
whose correlation matrix has the symbol 
\be\label{SM:ini}
\hat\Gamma_{x,0}(e^{i p})=-\theta_H(-x)e^{\eta x}\tanh\Bigl(\frac{\beta \hat h(e^{i p})}{2}\Bigr)\, ,
\ee
with $\beta=1$.
The initial state of Fig.~\ref{fig:1} corresponds to $\eta=0$. For technical reasons related to the numerical evaluation of \eqref{eq:exactsolhom}, we set $\eta=0.001$. The effect of a nonzero $\eta$ is to make the density matrix proportional to the identity in the limit $|x|\rightarrow\infty$ but, essentially, \eqref{SM:ini} describes the junction of two thermal states, the second of which has the temperature equal to infinity.  In the left panel of Fig.~\ref{fig:2}, the prediction \eqref{eq:exactsolhom}, specialized to the local magnetization $\braket{\boldsymbol s^{z}_\ell}$ with $\ell=4$, is compared with  exact numerical data obtained in a chain with $201$ spins. The data are practically indistinguishable from the prediction, and the imperceptible discrepancy is due to to finite-size effects (the prediction is valid in the thermodynamic limit) and to errors in the evaluation of \eqref{eq:exactsolhom} (in order to speed up the numerical evaluation of the integral, we have introduced a cutoff). 

\begin{figure}[h]
\includegraphics[width=0.45\textwidth]{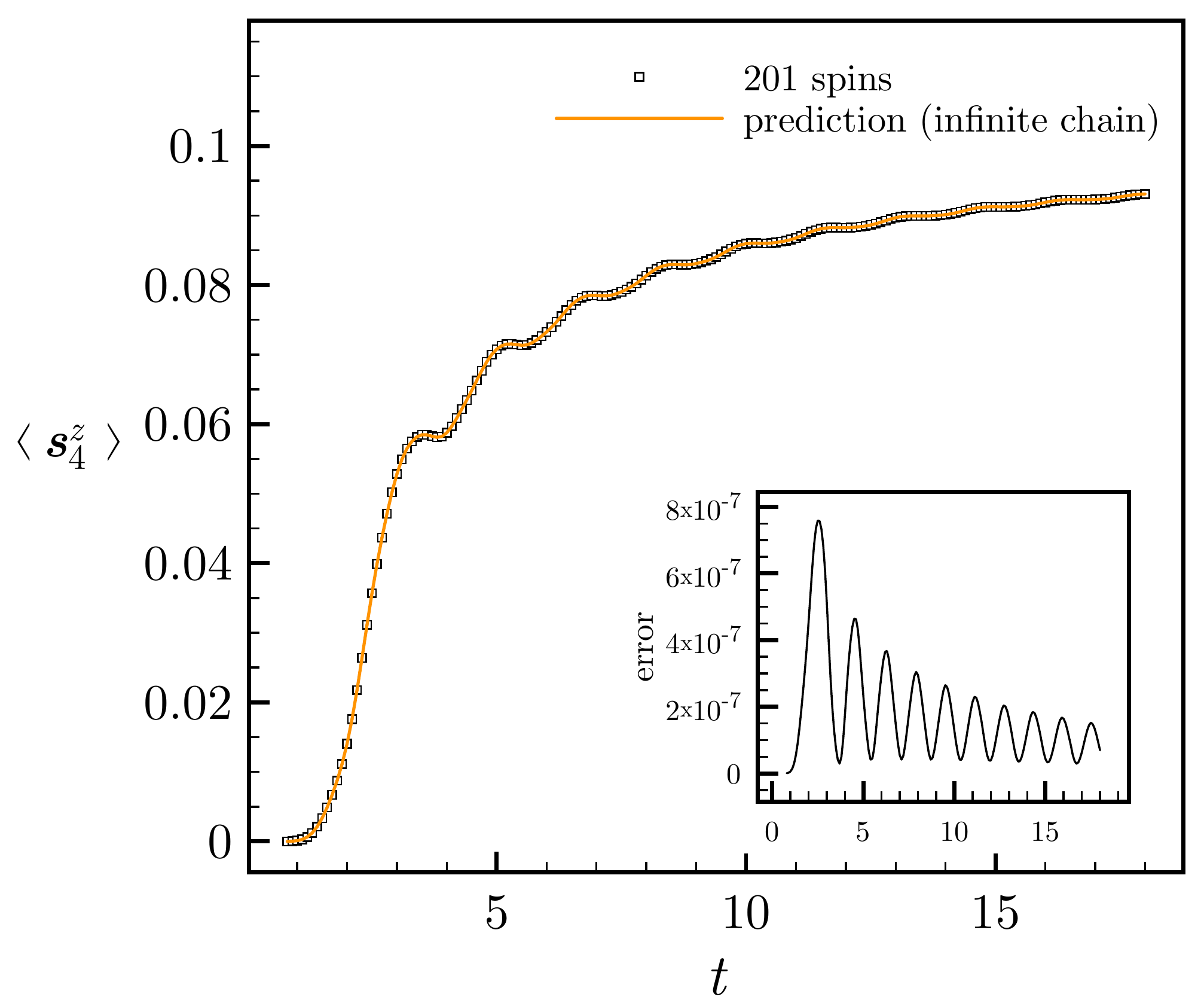}\qquad
\includegraphics[width=0.45\textwidth]{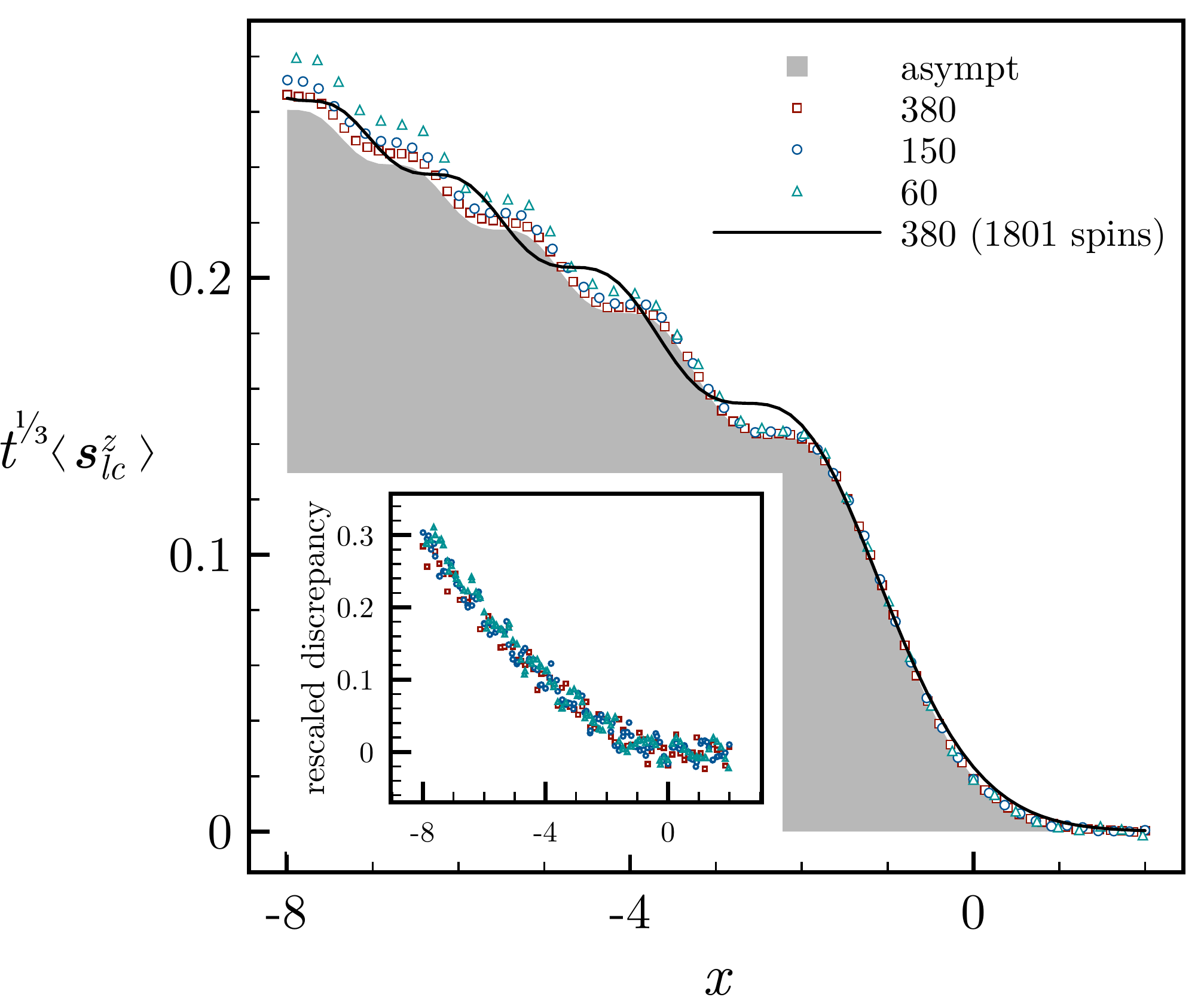}
\begin{center}
\caption{\emph{Left panel}. The prediction \eqref{eq:exactsolhom} applied to the transverse magnetization compared with exact numerical data obtained in a chain with $201$ spins. The initial state is described by \eqref{SM:ini} with $\beta=1$ and $\eta=0.001$;  time evolution is generated by the Hamiltonian \eqref{SM:HXY}. The agreement between the analytical prediction and the numerical data is perfect. The inset shows the difference between the prediction and the numerical data; the error is almost entirely due to the numerical evaluation of \eqref{eq:exactsolhom}.\\
\emph{Right panel}.  The local magnetization close to the right light cone. The (upper) boundary of the filled region is the asymptotic prediction~\eqref{eq:szlc}. The symbols are the third-order hydrodynamic solution for three different times: $60$, $150$, and $380$. The solid line corresponds to time evolution in a chain with 1801 spins at the time $t=380$. 
Despite the time being rather large, the numerical data are not consistent with~\eqref{eq:szlc}. 
The inset displays the rescaled discrepancy $t(\braket{\boldsymbol s^z_{lc}}^{\rm hyd_3}-\braket{\boldsymbol s^z_{lc}}^{\rm asympt})$ between the hydrodynamic solution and the asymptotic prediction; the points collapse to the same curve, showing that the hydrodynamic solution approaches \eqref{eq:szlc} as $t^{-1}$. 
}\label{fig:2}
\end{center}
\end{figure}

\item[Check of the scaling behavior close the light cone]
Here we check the validity of \eqref{eq:Gammasol}. 
For this test, we set $\eta=0$ in \eqref{SM:ini}, so we are actually considering the same initial state as in Fig.~\ref{fig:1}. 
We focus on the scaling behavior close to the right light cone. 
From \eqref{eq:Gammasol}, it follows that the local magnetization behaves as
\be\label{eq:szlc}
\braket{\boldsymbol s^z_{lc}}(x)\sim 2\pi \cos\bar \theta\tanh(\frac{\beta\bar \varepsilon}{2}) K_\alpha(x,x)\sqrt[3]{\frac{2\alpha}{-\bar v^{\prime\prime} t}}\, ,
\ee
where $x$ is the rescaled position, as defined in \eqref{eq:rescaled}.
In the right panel of Fig.~\ref{fig:2}, \eqref{eq:szlc} is compared with the third-order hydrodynamic prediction, which is obtained using \eqref{eq:hydro3sol}.  The data approach \eqref{eq:szlc} as $t^{-1}$, showing that the asymptotic expansion is correct. We also display the magnetization computed following the exact time evolution in a chain with 1801 spins. As  for the connected two-point function reported in Fig.~\ref{fig:1}, the numerical data do not seem to approach the hydrodynamic prediction.

We point out that the same analysis in the XX model and in the transverse-field Ising chain discloses an excellent agreement between hydrodynamics and exact time evolution. We also checked other models with $\alpha=1$ (also with next-nearest-neighbor couplings); we have always found that hydrodynamics prefectly describes the scaling behavior close to the light cone. 

\end{description}

\begin{thebibliography}{99}
\bibitem{f:3} Contrary to standard conventions, we call $\mathcal H$ a matrix despite its indices being not strictly positive.
\bibitem{f:4} 
Eq.~\eqref{eq:hydrostandard} follows from \eqref{eq:hydro} using that the symbol of a Hamiltonian with nearest neighbor couplings has the form $\hat h(e^{i p})=\hat A+\hat B\cos p+\hat C \sin p$.
\bibitem{f:1} The interpretation is not exact because the state is inhomogeneous~\cite{LONG}. This has counterintuitive consequences, starting from the fact that the root densities can be negative, exactly as the Wigner quasi-probability distributions in quantum mechanics.
\bibitem{f:2} In some cases (not in the examples considered in this paper), a state with the root density \eqref{eq:gensolhydro} could be (slightly) unphysical: a pair of eigenvalues of the corresponding correlation matrix exceeds $1$ (in absolute value). This is an effect of the inhomogeneity, which could be cured by a tiny modification of the initial root density~\cite{LONG}.
\end{thebibliography}
\end{document}